\documentclass{article}
\usepackage{spconf,amsmath,graphicx}
\usepackage{graphicx,wrapfig}
\usepackage{subfig}

\title{Distributed Optimization via Adaptive Regularization for Large Problems with Separable Constraints \thanks{\footnotesize \MakeLowercase{\uppercase{T}his work was supported by the \uppercase{D}epartment of \uppercase{D}efense under the \uppercase{AFOSR} \uppercase{G}rant \uppercase{FA9550-11-1-0210}, and by \uppercase{N}ational \uppercase{S}cience \uppercase{F}oundation under the \uppercase{NSF} \uppercase{G}rants \uppercase{CCF-1014908} and \uppercase{CCF-0963742.}} }\vspace{-0.3cm}}
\name{Elad Gilboa, Phani Chavali$^*$, Peng Yang$^*$\thanks{$^*$ \footnotesize Equally contributing authors} and Arye Nehorai \vspace{-0.3cm}}
\address{\{gilboae,chavalis, yangp, nehorai\}@ese.wustl.edu\\Preston M. Green Department of Electrical and Systems Engineering \\Washington University in St. Louis,  St. Louis, MO 63130}

\newcommand{\bsg}{\boldsymbol{g}}
\newcommand{\bsd}{\boldsymbol{d}}

\newcommand{\bfP}{\boldsymbol{P}}

\newcommand{\bsh}{\boldsymbol{h}}

\newcommand{\bsx}{\boldsymbol{x}}

\newcommand{\bsy}{\boldsymbol{y}}

\newcommand{\mathbb}{\rm}

\newcommand{\gv}[1]{\ensuremath{\mbox{\boldmath$ #1 $}}}
\newcommand{\grad}[1]{\gv{\nabla} #1} % for gradient
\newcommand {\argmin}[1]{{\arg \mathop{\min}\limits_{#1}}}

\begin{document}
\maketitle
\begin{abstract}
\vspace{-0.1cm}
Many practical applications require solving an optimization over large and high-dimensional data sets, which makes these problems hard to solve and prohibitively time consuming. In this paper, we propose a parallel distributed algorithm that uses an adaptive regularizer (PDAR) to solve a joint optimization problem with separable constraints. The regularizer is adaptive and depends on the step size between iterations and the iteration number.
%With the regularization introduced, the joint problem can be partitioned into several sub-problems, each of which can be solved in parallel.
We show theoretical converge of our algorithm to an optimal solution, and use a multi-agent three-bin resource allocation example to illustrate the effectiveness of the proposed algorithm. Numerical simulations show that our algorithm converges to the same optimal solution as other distributed methods, with significantly reduced computational time.
\end{abstract}
%

%\begin{keywords}
%distributed optimization, multi-agent optimization, regularization, adaptive penalty, parallel update
%\end{keywords}
\vspace{-0.3cm}
\section{Introduction}
\vspace{-0.3cm}
%%%%%%%%%%%%%%%%% IMPORTANCE OF PARALLEL
%We develop a distributed method to solve optimization problems with large data sets of the following type:
%\begin{eqnarray}
%\label{jointminimization}
%\mbox{minimize} &&f(\bsx)\\
%\mbox{subject to} &&\bsx\in\mathcal{X},
%\end{eqnarray}
%where $\bsx \in \mathbb{R}^{d}$, where $d$ is very large.
With the sensor and the storage technologies becoming increasingly cheaper, modern applications are seeing a sharp increase in \emph{big data}. The explosion of such high-dimensional and complex data sets makes optimization problems extremely hard and prohibitively time consuming \cite{Boyd:2010}. Parallel computing has received a significant attention lately as an effective tool to achieve the high throughput processing speeds required for processing big data sets. Thus, there has a been a paradigm shift from aggregating multi-core processors to utilizing them efficiently \cite{Bradley:2011}.

Although distributed optimization has been an increasingly important topic, it has not received sufficient attention since the seminal work by Bertsekas and Tsitsiklis until recently. In the 1980's, Bertsekas and Tsitsiklis extensively studied decentralized detection and consensus problems \cite{Bertsekas:1989} and developed algorithms such as parallel coordinate descent \cite{Tsitsiklis:1986} and the block coordinate descent (BCD) (also called the block Jacobi) \cite{Bertsekas:1989,Tseng:2001}. In 1994, Ferris \emph{et.\ al.} proposed parallel variable distribution (PVD) \cite{Ferris:1993} that alternates between a parallelization and a synchronization step. In the parallelization step, several sub-optimal points are found using parallel optimizations. Then, in the synchronization step, the optimal point is computed by taking an optimal weighted average of the points found in the parallel step. Although PVD claims to achieve better convergence rate than BCD, the complexity of solving optimization in both the steps make it impractical for high dimensional problems. There are other efficient distributive methods in literature, such as the shooting \cite{Fu:1998}, the shotgun \cite{Bradley:2011}, and the alternating direction method of multipliers (ADMM) \cite{Boyd:2010}, however, these methods apply to only a specific type of optimization problems: $\ell_1$-regularization for shooting and shotgun, and linear constraints for ADMM.

In this paper, we propose a fully distributed parallel method to solve optimization problems over high-dimensional data sets, which we call the parallel distributive adaptive regularization (PDAR). Our method can be applied to a wide variety of nonlinear problems where the constraints are block separable. The assumption of block separable constraints holds good in a lot of practical problems, and can be commonly seen in problems such as multi-agent resource allocation. In order to coordinate among the subproblems we introduce an adaptive regularizer term that penalizes the large changes in successive iterations. Our method can be seen as an extension of the classical proximal point method (PPM) \cite{Bertsekas:1995} with two novel advances. First, our motivation for using the PPM framework is very different than the original. We use PPM as a means to coordinate among the parallel subproblems and not for handling non differentiability. Second, we enforce coordination by using adaptive regularizers that vary across different subproblems.

The rest of the paper is organized as follows. In Section\ \ref{sec:formulation}, we formulate the problem; in Section\ \ref{sec:alg} we propose our parallel distributive algorithm and show converges to an optimum solution; in Section\ \ref{sec:sims} we provide numerical simulations, and we conclude the paper in Section\ \ref{sec:conclude}.

\section{Problem Formulation}
\label{sec:formulation}
Consider an optimization problem given as:
\begin{eqnarray}
\label{jointminimization}
\mbox{minimize} &&f(\bsx)\\
\mbox{subject to} &&\bsx\in\mathcal{X},
\end{eqnarray}
where the objective is to find the optimal vector $\bsx^*$ that minimizes the function $f(\bsx)\in \mathbb{R}$, with $\bsx \in {\mathbb{R}}^d$. The problem is often very complex, nonlinear, and high dimensional, and solving it is prohibitively time consuming. We assume that the constraint $\bsx \in \mathcal{X}$ can be separated into several blocks, such that
\begin{eqnarray}
\label{unionatkp1}
\bsx = \left[\bsx_{1},  \bsx_{2},  \ldots, \bsx_{i},  \ldots,  \bsx_{N} \right] \ \text{where}, \ \bsx_{i} \in \mathcal{X}_i,
\end{eqnarray}
with $\bsx_{i} \in \mathbb{R}^{n_i}$ and $\sum_{i=1}^N n_i = d$. Once the problem is separated into blocks, distributed iterative approaches (such as the ones mentioned in the Introduction section) can be applied. However, these methods are time consuming when the sub-problems are themselves complex.

\section{Distributed Optimization via Adaptive Regularization}
\label{sec:alg}
In this section, we describe our distributed optimization framework with adaptive regularization. We solve the optimization problem given by Eq.\ (\ref{jointminimization}) in a parallel and iterative manner. Let $k$ denote the iteration index and $\hat{\bsx}^{k} = (\hat{\bsx}_{i}^k, \hat{\bsx}_{-i}^k)$, with $\hat{\bsx}_{-i}^k = \left[\hat{\bsx}_{1}^k, \ldots, \hat{\bsx}^k_{i-1}, \hat{\bsx}^k_{i+1}, \ldots, \hat{\bsx}^k_{N} \right]$ denote the solution to the optimization problem in the $k^{th}$ iteration. In order to obtain a solution in a distributed manner, we define a set of $N$ augmented objective functions at each iteration $k$ as  \footnote{We use a semicolon notation in Eq.\ (\ref{augmentedObjective}) to clarify that only the variables on the left of the semicolon are allowed to change.}
\begin{eqnarray}
\label{augmentedObjective}
L^{k}_i(\bsx_{i}; \hat{\bsx}^{k-1}) = f(\bsx_{i}, \hat{\bsx}^{k-1}_{-i})+{\lambda}^{k}_i(h_i^{k-1})\|\bsx_{i}-\hat{\bsx}^{k-1}_{i}\|^2,
\end{eqnarray}
where $\bsh^{k-1}_i = \hat{\bsx}_{i}^{k-1}-\hat{\bsx}^{k-2}_{i}$ is the step taken by the $i^{th}$ block in the $(k-1)^{th}$ iteration, and ${\lambda}^{k}_i(h_i^{k-1})$ is an adaptive regularization coefficient which depends on both the indices $i$ and $k$. We will describe the form of this regularization coefficient shortly. After defining the objective functions $L^{k}_i(\raisebox{1mm}{.}), \ i = 1, \ldots, N$, we solve $N$ optimization problems in a parallel fashion:
\begin{eqnarray}
\label{ParallelBlockRelaxationMinimization}
\hat{\bsx}_{1}^{k} &=& \argmin{\bsx_{1}\in\mathcal{X}_1} L^{k}_i(\bsx_{1}; \hat{\bsx}^{k-1}), \nonumber \\
\hat{\bsx}_{2}^{k} &=& \argmin{\bsx_{2}\in\mathcal{X}_2} L^{k}_i(\bsx_{2}; \hat{\bsx}^{k-1}), \nonumber \\
&\vdots& \nonumber \\
\hat{\bsx}_{N}^{k} &=& \argmin{\bsx_{N}\in\mathcal{X}_N} L^{k}_i(\bsx_{N}; \hat{\bsx}^{k-1}).
\end{eqnarray}
This optimization framework is in the form of a decomposition-coordination procedure \cite{Boyd:2010}, where $N$ agents are trying to minimize their own augmented objective functions, and the new joint vector $\hat{\bsx}^{k}$ is obtained by simply aggregating the $N$ blocks. Further, since the minimization of the objective functions $L^{k}_i(\raisebox{1mm}{.})$ is only with respect to the variables of the $i^{th}$ block and the other blocks are constants which change with every iteration, the objective functions will change in every iteration.

Next, we discuss the choice of the regularization coefficient ${\lambda}^{k}_i(h_i^{k-1})$. We chose ${\lambda}^{k}_i(h_i^{k-1})$ to be of the form:

\begin{eqnarray}
\label{lambdafunction}
\lambda_i^k(h^{k-1}_i) = \left\{ \begin{array}{ll}
\max (\phi(\|h^{k-1}_i \|),\beta) & \mbox{if $k < K$} \\
\alpha k & \mbox{otherwise,} \\
\end{array}
\right.
\end{eqnarray}
where $K$ is a threshold on the iteration index, $\alpha > 0$, and $\beta > 0$ are parameters chosen depending on the problem. Intuitively, the threshold $K$ divides each optimization problem into two phases. The goal of the first phase is to coordinate the parallel optimization. In this phase, each of the agents change their solution in response to the solutions of other agents. This alternating behavior can be enforced by choosing the function $\phi(\|h^{k-1}_i \|)$ to be a nondecreasing with respect to $\|h^{k-1}_i\|$. This choice will increase the value of regularization coefficient, $\lambda_i^k(h^{k-1}_i)$ as $\|h^{k-1}_i \|$ increases. The increase in $\lambda_i^k(h^{k-1}_i)$ will in turn enforce a smaller stepsize on the agents that had large change in the previous iteration, to allow other agents to react in the current iteration. The goal of the second phase is to fine tune the solution and to enable it reach a local optimum.  Although the choice of the function $\phi$ and the parameters $\alpha$, and $\beta$ theoretically  effect the convergence of the optimization, we observed using numerical simulations that the convergence was not sensitive to these choices. In this paper we choose $\phi(\|h^{k-1}_i \|) = N^2 \|h^{k-1}_i \|$.
%We also mention here that choosing a non-adaptive regularizer $\phi(\|h^{k-1}_i \|) = C$, will make it hard to select $C$ that will ensure that all the subproblems converge.
The algorithm is summarized in Table\ \ref{tab:distopt}.

\begin{table}
\begin{center}
\small
\begin{tabular} {l} \hline
\textbf{Algorithm: PDAR} \\ \hline
$k = 1$;  \% Iteration counter \\
Initialize $\bsx^{0}$ and $\lambda_{i}^0 \ \forall \ i$ \\
\textbf{do} \\
\qquad \textbf{parfor $i$ in $1:N$} \\
\qquad \qquad $\hat{\bsx}_{i}^{k} = \argmin{\bsx_{i}\in\mathcal{X}_i} L^{k}_i(\bsx_{i}; \hat{\bsx}^{k-1})$ \\
\qquad \qquad Set $\bsh^{k}_i = \hat{\bsx}_{i}^{k}-\hat{\bsx}^{k-1}_i$ \\
\qquad \qquad Update $\lambda_{i}^k$ \\
 \qquad \textbf{end parfor} \\
\qquad $k := k+1$ \\
\textbf{until} \footnotesize{ $\|f(\bsx^{k})-f(\bsx^{k-1})\| \leq \delta$ } \\
\hline  \hline
\end{tabular}
\end{center}
\caption {Algorithm for Parallel Distributed Optimization}
\label{tab:distopt}
\end{table}

\subsection{Discussion on the Convergence}

In this section, we show that the algorithm described in the previous subsection converges to an optimum solution. Assume that the function $f(\bsx)$ is convex. Since the augmented function $L^{k}_i(\raisebox{1mm}{.})$, $i = 1, \ldots, N$ is the sum of two convex functions,  it is convex.  We then have
\begin{eqnarray}
\label{simpleSubproblem}
\hat{\bsx}_{i}^{k} &=& \argmin{\bsx_{i}\in\mathcal{X}_i} L^{k}_i(\bsx_{i}; \hat{\bsx}^{k-1}).
\end{eqnarray}
%Constructing a nonnegative penalty function $\lambda$ results in
%\begin{eqnarray}
%\label{fxkBiggerThanfxkp1i}
%f(\bsx^k) = z_{\bsx^k}(\bsx^{k}) &\geq& z_{\bsx^k}(\hat{\bsx}^{k+1}_i)  = f(\hat{\bsx}^{k+1}_i)+\underbrace{\lambda_i(k,h^{k-1}_i)\|\hat{\bsx}^{k+1}_i-\bsx^k\|^2}_{\geq 0} \geq f(\hat{\bsx}^{k+1}_i).
%\end{eqnarray}
Since $\hat{\bsx}_{i}^{k}$ is a minimizer of $L^{k}_i(\bsx_{i}; \hat{\bsx}^{k-1})$, we have by the first order necessary conditions for local optimum that
\begin{eqnarray}
\label{partialGradOfz}
&&\grad_i L^{k}_i(\bsx_{i}; \hat{\bsx}^{k-1})\bigg|_{\bsx_{i} = \hat{\bsx}_{i}^{k}} = 0, \nonumber  \\
&&\grad_i f(\hat{\bsx}^k_{i},\hat{x}_{-i}^{k-1}) + 2\lambda_i^{k}(h^{k-1}_i)\underbrace{(\hat{\bsx}^k_{i}-\hat{\bsx}^{k-1})}_{\bsh_i^k} = 0,\nonumber \\
&\Rightarrow& \grad_i f(\hat{\bsx}^k_{i},\hat{x}_{-i}^{k-1}) = -2\lambda_i^{k}(h^{k-1}_i)\bsh_i^k,\nonumber \\
\label{kp1StepSize}
&\Rightarrow& \bsh^{k}_i = \frac{- \grad_i f(\hat{\bsx}^{k}_{i},\hat{\bsx}^{k-1}_{-i})}{2\lambda_i^k(h^{k-1}_i)},
%&\Rightarrow& \bsh^{k}_i   = \frac{- \grad_i f(\hat{\bsx}^{k}_{i},\hat{\bsx}^{k-1}_{-i})}{2\max (\gamma \phi(\|h^{k-1}_i \|),\alpha k)},
\end{eqnarray}
where the operator $\grad_i$ is a gradient operator with respect to $\bsx_{i}$.
%From Eq.\ (\ref{partialGradOfz}) we can observe that at the point $\bsx = \bsx^k$ we have $\grad_i z_{\bsx^k}(\bsx^k_i)=\grad_i f(\bsx^k_i)$. Hence, if $\bsx^*$ is a local optimum for $f(\bsx)$, then $\forall i, ~ \grad_i f(\bsx_i) = 0$, so that $\bsx^*$ is also optimal for $z_{\bsx^*}(\bsx)$.
%Assuming that the step size is bounded, and putting a bound on $\phi$, then $\exists M_1,M_2 \in \mathbb{R},~ s.t.$
%\begin{eqnarray}
%\forall i,~ \|h^{k-1}_i \| \leq M_1,~ \phi(\|h^{k-1}_i \|) \leq M_2.
%\end{eqnarray}
%\label{lambdaForLargeK}
For $k > K$ we have $ \lambda_i^k(h^{k-1}_i) = \alpha k$, and therefore Eq.\ (\ref{kp1StepSize}) simplifies as
\begin{eqnarray}
\bsh^{k}_i = \frac{1}{2 \alpha k} \underbrace{\left(- \grad_i f(\hat{\bsx}^{k}_{i},\hat{\bsx}^{k-1}_{-i})\right)}_{\bsd_i^k},
\end{eqnarray}
where $\bsd_i^k$ is the negative gradient direction of the $i^{th}$ agent. By concatenating all the directions into a single vector $\bsd^k = [\bsd^{k}_1, \bsd^{k}_2, \ldots, \bsd^{k}_N]$,  we get the next iterate ${\bsx}^{k}$ as
\begin{eqnarray}
\hat{\bsx}^{k} &=& \hat{\bsx}^{k-1}+\bsh^{k},
\end{eqnarray}
where $\bsh^{k} = \frac{\bsd^k}{2 \alpha k}$.  We prove the convergence properties of the algorithm using the following two prepositions. \\
%Formally, we need to show that for any subsequence $\{x^k\}_{k\in K}$ that converges to a nonstationary point, the corresponding subsequence $\{\bsd^{k}\}$ is bounded and satisfies
%\begin{eqnarray}
%\label{gradeintRelated}
%\lim_{k\rightarrow \infty} {\rm{sup}}_{k\in \mathcal{K}} \grad f(\bsx^k)'d(\bsx^k) <0.
%\end{eqnarray}

\noindent\textbf{Proposition 1:} For the sequence of non-stationary iterates $\hat{\bsx}^{k}$ obtained from the PDAR algorithm, $\grad f(\hat{\bsx}^{k-1})' \bsd^{k} <0$. \footnote{For brevity, if all the blocks in the function are from the same iteration, we will simplify the notation, i.e., $f(\hat{\bsx}_i^{k-1},\hat{\bsx}_{-i}^{k-1}) = f(\hat{\bsx}^{k-1})$} \\
\emph{Proof:} From the definition of $\bsd^k_i$, we have
\begin{equation}
\bsd^k_i = - \grad_i f(\hat{\bsx}^{k}_{i},\hat{\bsx}^{k-1}_{-i}).
\end{equation}
Therefore,
\begin{eqnarray}
\label{dotprocdef}
\grad f(\hat{\bsx}^{k-1})' \bsd^{k} = \sum_{i=1}^N - \grad_i f(\hat{\bsx}^{k-1}) \grad_i f(\hat{\bsx}^{k}_{i},\hat{\bsx}^{k-1}_{-i}).
\end{eqnarray}
Since $\hat{\bsx}^{k}_i$ is a result of minimizing $L^{k}_i(\bsx_{i}; \hat{\bsx}^{k-1})$,
the corresponding step $\bsh^{k}_i$ must be in a descending direction. Thus
\begin{eqnarray}
\grad_i  L^{k}_i(\hat{\bsx}^{k-1})'\bsh^{k}_i = \grad_i f(\hat{\bsx}^{k-1})' \bsh^{k}_i \leq 0, \ \  \forall \ i
\end{eqnarray}

However, there must exist at least one block where the strict inequality $\grad_i f({\bsx}^{k-1}_i)' \bsh^{k}_i < 0$ holds. We prove this by contradiction. Assume that $\forall i,~\grad_i f({\bsx}^{k-1}_i)' \bsh^{k}_i = 0$. If $\bsh^{k}_i = 0, \forall i$, then $\hat{\bsx}^k$ is a stationary point which contradicts the assumption of convergence to a nonstationary point. Hence there exists some $i$, for which $\bsh^{k}_i\neq 0$. Now, since $L^k_i({\bsx}^k; \hat{\bsx}^{k-1})$ is a convex function, it must lie above all of its tangents, i.e.,
\begin{eqnarray}
\label{convexAboveTangents}
L^k_i(\hat{\bsx}_i^{k};\hat{\bsx}_{-i}^{k-1}) \geq L^k_i(\hat{\bsx}^{k-1})+ \grad_i L^k_i(\hat{\bsx}^{k-1})' h^k_i.
\end{eqnarray}
Since $\grad_i  L^k_i(\hat{\bsx}^{k-1})' \bsh^{k}_i = \grad_i f(\hat{\bsx}^{k-1})' \bsh^{k}_i = 0$, we have from Eq.\ (\ref{convexAboveTangents}), that $L^k_i(\hat{\bsx}^{k}) \geq L^k_i(\hat{\bsx}^{k-1})$. This is a contradiction, since every iterate should reduce the objective function corresponding to the block.  Intuitively, this inequality implies that if the step size is perpendicular to the gradient of the objective function, then such steps do not decrease the value of the objective function. Hence there exists at least one block that satisfies inequality $\grad_i f({\bsx}^{k-1}_i)' \bsh^{k}_i < 0$. Finally, since at least one block satisfies the strict inequality, their summation satisfies strict inequality:
\begin{eqnarray}
&&\sum_{i=1}^N \grad_i f(\hat{\bsx}^{k-1}_i)' \bsh^{k}_i < 0, \nonumber \\
&\Rightarrow& \sum_{i=1}^N \grad_i f(\hat{\bsx}^{k-1})' \grad_i f(\hat{\bsx}^{k}_{i},\hat{\bsx}^{k-1}_{-i}) < 0, \nonumber \\
%\sum_{i=1}^N - 2 \underbrace{\lambda_i(k,h^{k-1}_i)}_{>0}\grad_i f(\bsx^k)' \bsh^{k}_i > 0\\
\label{fd_ineq}
&\Rightarrow & \grad f(\hat{\bsx}^{k-1})' \bsd^{k} < 0. \nonumber
\end{eqnarray}

\noindent\textbf{Proposition 2:} The sequence $\hat{\bsx}^{k}$ converges to an optimal solution.  \\
\emph{Proof:} Assume that $f$ satisfies Lipschitz continuity of the gradients, and that its gradients are bounded. Formally, we need to show that for any subsequence $\{\hat{\bsx}^k\}$ that converges to a nonstationary point, the corresponding subsequence $\{\bsd^{k}\}$ is bounded and satisfies \cite{Bertsekas:1995}:
\begin{eqnarray}
\label{gradeintRelated}
\lim_{k\rightarrow \infty} {\rm{sup}}_{k \in \mathcal{K}} \grad f(\hat{\bsx}^{k-1})'\bsd(\hat{\bsx}^k)< 0,
\end{eqnarray}
where $ \bsd(\hat{\bsx}^k)= -\sum_{i=1}^N \grad_i f(\hat{\bsx}^k_{i}, \hat{\bsx}^{k-1}_{-i})$. Let $\epsilon>0$, and $\{\bsx^k\}_{k\in \mathcal{K}}$ be an arbitrary sequence of nonstationary points such that
\[\lim_{k \rightarrow \infty} {\rm{sup}}_{k \in \mathcal{K}} \hat{\bsx}^k = \bar{{\bsx}},\]
where $\grad f(\bar{\bsx}) \ne 0$. Then $\forall k \in \mathcal{K}$ the gradients are not equal to zero, $\grad f({\hat{\bsx}^k}) \ne 0$, since the sequence has nonstationary points. Using Proposition $1$, we have that $\forall k \in \mathcal{K}, ~\grad f({\hat{\bsx}^{k-1}})' d(\bsx^k) <0$, and specifically $\grad f(\bar{\bsx})' d(\bar{\bsx})=D_1 <0$.

By the Lipschitz continuity assumption of the gradients, there $\exists \ \delta>0$ such that $\| \grad f(\bsy)'d(\bsy)-\grad f(\bar{\bsx})'d(\bar{\bsx})\| < \epsilon$,  $\forall \ \|\bsy - \bar{\bsx}\|<\delta$.
Since $\bsx^k \rightarrow \bar{\bsx}$, $\exists \ N \ \in \mathbb{N}$ such that $\forall k >N,~  \|\bsx^k-\bar{\bsx}\|<\delta$, and thus
\[\| \grad f(\bsx^{k-1})'d(\bsx^k)-\grad f(\bar{\bsx})'d(\bar{\bsx})\| < \epsilon.\]
This implies that $\grad f(\bsx^{k-1})'d(\bsx^k)<D_1+\epsilon$. As $\epsilon>0$ is arbitrary, $\lim_{k\rightarrow \infty} {\rm{sup}}_{k\in \mathcal{K}} \grad f(\bsx^{k-1})'d(\bsx^k)=D_1 <0$. Hence the sequence of iterates $\bsx^k$ converges to an optimal solution.

\section{Numerical Results}
\label{sec:sims}
In this section, we provide numerical results to compare the convergence of the proposed distributed algorithm to those of the block coordinate descent (BCD) and parallel variable distribution (PVD) . We consider a three-bin resource allocation example for the numerical simulation. Let there be $N = 100$ agents. Each agent has fixed quantity of resources that are to be allocated among three bins. Let $\bsx_{i} = [x_{i,1}, x_{i,2}, x_{i,3}]'$ denote the allocation scheme of the $i^{th}$ agent. Without loss of generality, let $\sum_{j=1}^{3}x_{i,j} = 1, \ \forall \ i$. The objective is to minimize the sum of the individual costs, where the cost of agents depends on their own scheme and the schemes of other agents.

Let $\bsx = [\bsx_1', \bsx_2', \ldots, \bsx_N']'$ denote the collective scheme of all agents. The cost function of the $i^{th}$ agent is taken as
\begin{eqnarray}
f_i(\bsx) = \bsx_{i}' \bfP_i \bsg(\bsx),
\end{eqnarray}
where $\bfP_i = \mathrm{diag} (p_{i,1}, p_{i,2},p_{i,2})$ denotes the preference matrix of the $i^{th}$ agent for each bin, and $\bsg(\bsx) = [g_1, g_2, g_3]'$ is a function dependent on the schemes of all agents, with
\begin{equation}
g_m = \left(\sum_{i=1}^N x_{i,m}\right)^2, \quad m \in\{1,2,3\}.
\end{equation}
The goal is to solve the optimization problem:
\begin{eqnarray}
\mathop{\min}_{\bsx} \sum_{i=1}^{N}f_i(\bsx) \ \ \mathrm{subject~to~} \sum_{j=1}^{3}x_{i,j} = 1, \forall \ i.
\end{eqnarray}
In order to find the solution to the above joint optimization problem, we solved $N=100$ subproblems in parallel using our proposed PDAR. The optimization problem of the $i^{th}$ agent in the $k^{th}$ iteration is given as

\begin{eqnarray}  \label{eqn:distoptpen}
\begin{aligned}
& \mathop{\min}_{\bsx_{i}} & & f_i(\bsx_{i},\hat{\bsx}^{k-1}_{-i}) + \lambda_i^k(h_i^{k-1})\|\bsx_{i}-\hat{\bsx}^{k-1}_{i}\|^2 \\
& \mathrm{subject~to~} & & \sum_{j=1}^{3}x_{i,j} = 1.
\end{aligned}
\end{eqnarray}

\begin{figure}[h]
  \centering
    \subfloat[]{\label{scrFun} \includegraphics[height = 0.25\textwidth, width=0.4\textwidth]{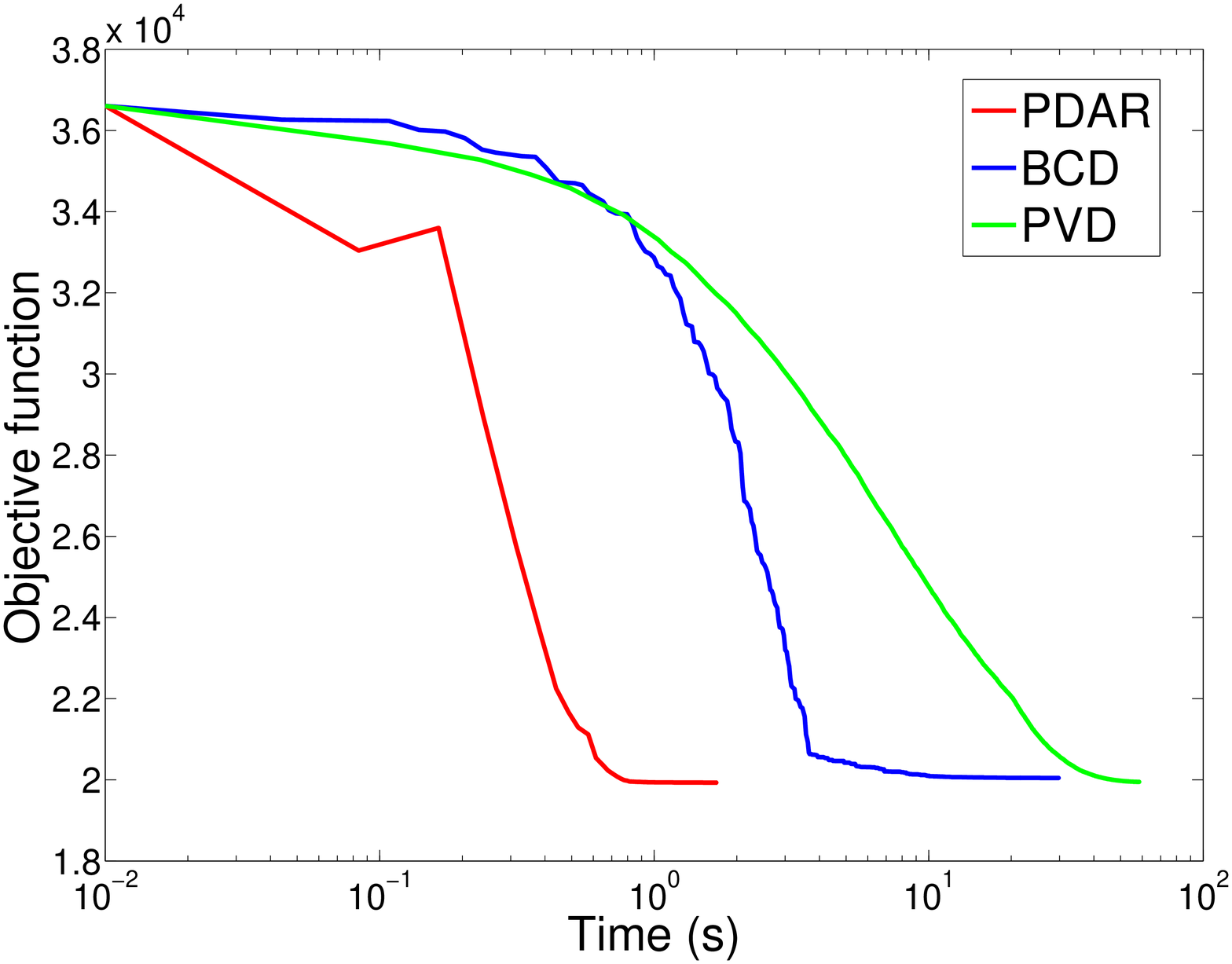}}\\
    \subfloat[]{\label{noreg} \includegraphics[height = 0.25\textwidth, width = 0.4\textwidth]{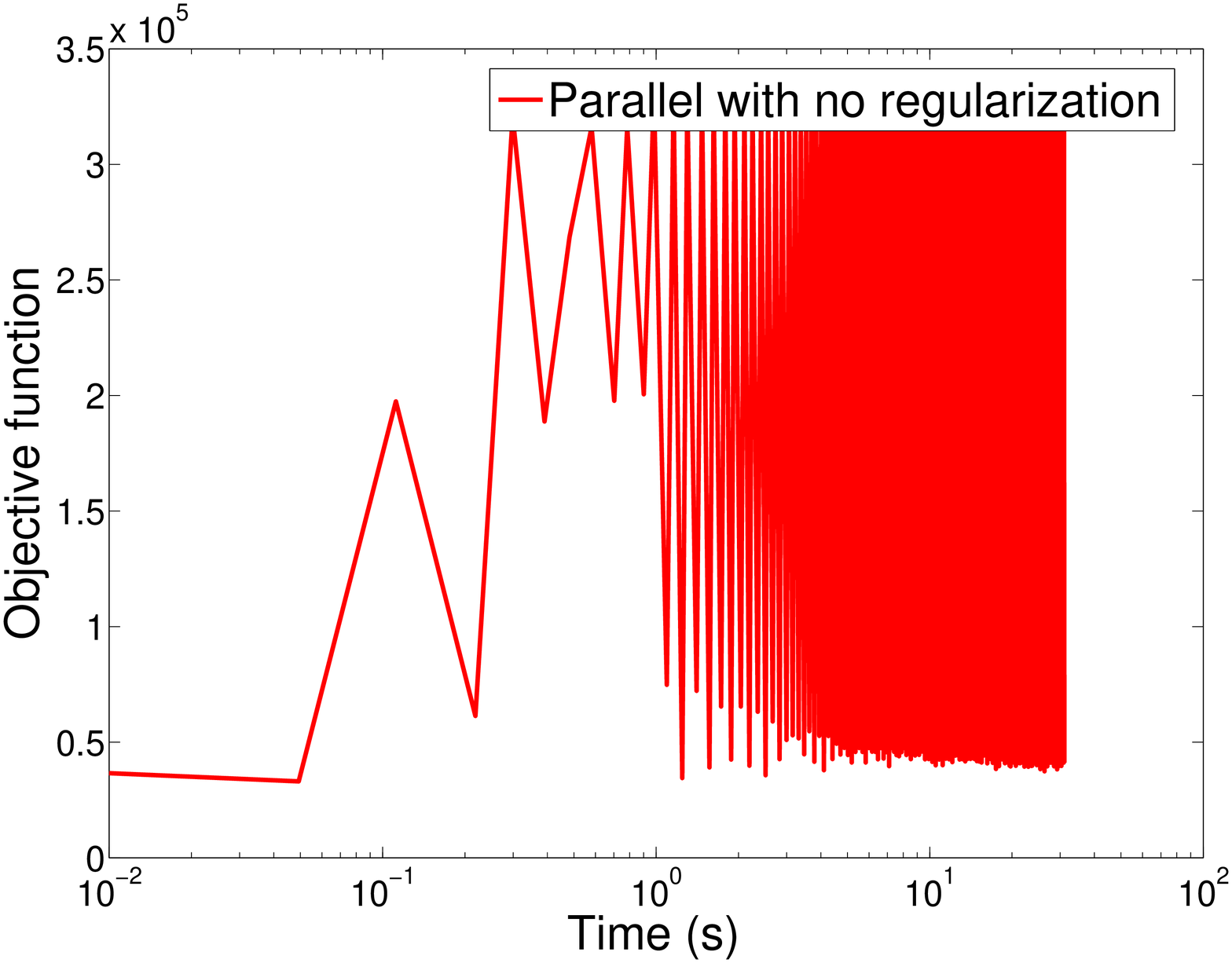}}
    \caption{Value of the objective functions vs time for the three bin resource allocation problem. Fig.\ \ref{scrFun} shows that PDAR converges much faster compared to BCD and PVD. Fig.\ \ref{noreg} shows the oscillatory behavior of the parallel optimization without regularization.}
    \label{fig:scoreFn}
\end{figure}

In Fig.\ \ref{scrFun}, we plot the value of the objective function as a function of the normalized time for BCD, PVD and our PDAR approach. We ran all the simulations on a 4 core machine. However in principle the parallel methods can be run on 100 cores simultaneously. Hence, it order to make the comparison fair, the time axis corresponding to parallel methods was divided by 25. As illustrated, the convergence rate of our method is of an order of magnitude faster compared to BCD and PVD algorithms. The advantage comes from the fact that we can solve all the $100$ optimization problems in parallel, whereas BCD is a sequential method. The PVD method, on the other hand, is worse even though it has a parallel update step. The additional time it takes to converge is due to the synchronization step, and  due to the complexity of the optimization problems that are to be solved in both steps. In Fig.\ \ref{noreg}, we show the oscillatory behavior when the parallel algorithm is used with out a regularizer. This figure further emphasizes the importance of a regularizer.

\section{Conclusions}
\label{sec:conclude}
In this paper, we proposed a distributed optimization framework to solve large optimization problems with separable constraints. Each agent solves a local optimization problem, which is much simpler compared to the joint optimization. In order for the agents to coordinate among themselves and to reach an optimum solution, we introduced a regularization term that penalized the changes in the successive iterations with an adaptive regularization coefficient. We proved that our solution always converges to a local optimum, and to a global optimum if the overall objective function is convex. Numerical simulations showed that the solutions reached by our algorithm are the same as the ones obtained using other distributed approaches, with significantly reduced computation time.

\bibliographystyle{IEEEtran}
\bibliography{ParallelBib}

\end{document}